\def\half{\frac{1}{2}}
\def\mb#1{\mbox{\boldmath{$#1$}}}
\def\sgn{{\rm sgn}}
\def\eq#1{Eq.\,(\ref{#1})}
\def\nl{\hfil\break\noindent}
\def\dbox#1{\hbox{\vrule  
                        \vbox{\hrule \vskip #1
                             \hbox{\hskip #1
                                 \vbox{\hsize=#1}%
                              \hskip #1}%
                         \vskip #1 \hrule}%
                      \vrule}}

\def\qed{\hfill \dbox{0.05true in}}  

\documentclass[amsmath,amssymb,superscriptaddress,showkeys, showpacs]{revtex4}
\usepackage{amsmath}
\usepackage{graphicx}
\usepackage{rotating}
\usepackage{mathrsfs} 
\begin{document}
\noindent\hspace*{4 in}CUQM-142, HEPHY-PUB 902/11\\
\def\ttitle{Geometric spectral inversion for singular potentials}
\title{\ttitle}
\markboth{R.~L.~Hall \& W.~Lucha}{\ttitle}

\author{Richard L. Hall}
\email{rhall@mathstat.concordia.ca}
\affiliation{Department of Mathematics and Statistics, Concordia University,
1455 de Maisonneuve Boulevard West, Montr\'eal,
Qu\'ebec, Canada H3G 1M8}
\author{Wolfgang~Lucha}
\email{wolfgang.lucha@oeaw.ac.at}
\affiliation{Institute for High Energy Physics, Austrian Academy
of Sciences, Nikolsdorfergasse 18,\\ A-1050 Vienna, Austria}

\begin{abstract}
The function $E = F(v)$ expresses the dependence of a discrete eigenvalue $E$ of the Schr\"odinger Hamiltonian $H = -\Delta + vf(r)$ on the coupling parameter $v$. We use envelope theory to generate a functional sequence $\{f^{[k]}(r)\}$ to reconstruct $f(r)$ from $F(v)$ starting from a seed potential $f^{[0]}(r).$ In the power-law or log cases the inversion can be effected analytically and  is complete in just two steps.  In other cases convergence is observed numerically.  To provide concrete  illustrations of the inversion method  it is first applied to the Hulth\'en potential, and it is then used to invert spectral data generated by singular potentials with shapes of the form $f(r) = -a/r + b\,\sgn(q)r^q$ and $f(r) = -a/r + b\ln(r),$ $a,\,b>0.$ For the class of attractive central potentials with shapes $f(r) = g(r)/r,$ with $g(0)< 0$ and $g'(r)\ge 0,$ we prove that the ground-state energy curve $F(v)$ determines $f(r)$ uniquely. 
\end{abstract}



\maketitle
\section{Introduction}

The purpose of this paper is to report a robust method for reconstructing the detailed shape of a singular attractive potential from some of the bound-state spectral data that it generates. The general setting for the problem  is the discrete spectrum of a Schr\"odinger Hamiltonian operator 
\begin{equation}\label{hamiltonian}
H = -\Delta +vf(r),\quad r\equiv \|\mb{r}\|,
\end{equation}
where $f(r)$ is the shape  of an attractive central potential, and $v >0$ is a coupling parameter.  We shall assume that the potential is monotone  non-decreasing and no more singular than the Coulomb potential $f(r) = -1/r.$ The arguments we use apply generally to the problem  in $d>1$ spatial dimensions, but, for definiteness, we shall usually assume that $d = 3.$ The operator  inequality \cite{GS,RS2} 
\begin{equation}\label{opineq}
-\Delta \ge \left(\frac{d/2-1}{r}\right)^2,\quad d \ge 3, 
\end{equation}
implies that a discrete spectrum exists for sufficiently large coupling $v >0$. For $d=3,$ the Hamiltonian $H$ is bounded below by
\begin{equation}\label{elower}
E \ge \min_{r > 0}\left[\frac{1}{4r^2} + vf(r)\right],
\end{equation}
and a simple trial function can be used to establish an  upper bound.
 Thus we may assume, in particular, that the ground-state energy may be written as a function $E = F(v)$. An explicit example of the class of problems we consider is provided by the Hulth\'en potential whose shape is given by $f(r) = -1/(e^{r}-1)$ and whose s-state ($\ell = 0$) eigenvalues $E_n$ are given \cite{flug} exactly for $d=3$ by the formula
\begin{equation}\label{ehulthen}
E_n = F_n(v) = -\left(\frac{v-n^2}{2n}\right)^2,\quad v > n^2,\quad n = 1,2,3,\dots.
\end{equation} 
The problem discussed  in the present paper may be stated as follows: given, for example, the curve $F_1(v),$ can we use this spectral data to reconstruct the potential shape $f(r)$? We call this `geometric spectral inversion'.  It must be distinguished from `inverse scattering theory' \cite{chad,newt,zak,eilen,toda,chada,chadb,chadc,chadd}, and,  in particular, from `the  inverse problem in the coupling parameter' \cite{chad}.  In this latter problem, the energy eigenvalue $E$ is fixed, and the data to be  inverted comprises the set of all possible couplings leading to a radial eigenstate with this given energy, along with certain scattering data. In geometric spectral inversion, we simply use one spectral curve, say $F_1(v).$ Progress has been reported on this problem in a sequence of earlier papers \cite{inv1,inv2,inv3,inv4,inv5}.  From the point of view  of problems in $d=1$ dimension, we were at first concerned that for excited states, which often only exist for $v$ sufficiently large, some essential pre-image data might be `missing'.  However, the analytical  inversion of the WKB approximation \cite{inv3} showed that the WKB inversion process improves in quality as one goes to higher states, and is moreover asymptotically exact. Progress has been made on a related problem, namely on inverting spectral data of the form $E(\ell)$, where $\ell$ is the orbital angular momentum quantum number, for fixed coupling $v;$ this work by Grosse and Martin is described in Ref. \cite{GM}.

We suggest two areas of application for the form of spectral inversion $F(v)\rightarrow f(r)$ which is the main topic of this paper.  The spectrum generated by the outermost electron in a large atom is often estimated by use of a screened-Coulomb potential with parameters dependent on the atomic number $Z$. It is envisaged that atomic spectral data for a sequence of atomic numbers could be used to reconstruct an appropriate effective potential for the model.  In another area, quantum field theory can be used to predict the dependence $E = F(v)$ of the energy of a two-body bound system on the coupling $v$.  Geometric spectral inversion could then be used to construct from $F(v)$ a potential $f(r)$ that would reproduce the same spectrum in a quantum-mechanical model. The idea is that instead of fitting the parameters in an appropriate potential model, we  instead develop an inversion method that allows us to reconstruct the potential shape directly from an eigenvalue curve.

In the present paper we study the inversion of spectral data arising from singular potentials.  We develop a functional inversion algorithm based  on envelope theory, a geometrical theory of energy functions such as $F_n(v).$ This was first introduced \cite{env1} in 1980 as a method for finding energy bounds for many-body problems, and has since led to a number of refinements and applications 
\cite{env2,env3,env4,env5,env6}, including semirelativistic problems \cite{halld,halle,hallf,hallg,hallh}. In 
sections~II and III we first review enough of this theory to serve our present purpose and to make this paper essentially
 self-contained.  In section~IV we develop  the inversion algorithm.  We then test the algorithm on some concrete problems. In section~V we show that the algorithm inverts spectral data arising from pure powers, or the log potential, exactly, in just two steps.  In section~VI we invert the Hulth\'en spectral data from Eq.\,(\ref{ehulthen}).  In section~VII  we apply the inversion algorithm to the spectral data generated by the family of  potentials $f(r) = -a/r +b\,w(r),$ where $w(r) = r,$ $w(r) = r^2,$ and $w(r) = \ln(r).$
In these latter problems, for which analytic expressions for the spectral curves $F(v)$ are not available, we begin in each case with the potential $f(r)$, find $F(v)$ numerically (by shooting methods), and then apply the sequential inversion method presented in this paper to reconstruct $f(r)$ again from $F(v).$ The spectral {\it inversion} $F\rightarrow f$ is the principal concern of the present work.  In section VIII we consider the ground  state energy curve $F(v)$ for the problem $H= -\Delta + vg(r)/r,$ where  $g(0) < 0$, and $g'(r) \ge 0.$ For this class of problems we prove that $F(v)$ determines $f(r) = g(r)/r$ uniquely.
\section{Exact representation of spectral functions by kinetic potentials}
The discrete spectra of operators such as $H = -\Delta + vf(r),$ which are bounded below, may be characterized variationally \cite{RS4}. Thus, the ground-state energy may be written
\begin{equation}\label{varchar}
F(v) =  \inf_{{{\scriptstyle \psi \in {\cal D}(H)} \atop {\scriptstyle \|\psi\| = 1}}} (\psi, H\psi).
\end{equation}
Since $H$ depends on the coupling $v,$ so therefore does the domain ${\cal D}(H).$ However, for the problems considered, either $H$ has discrete eigenvalues, perhaps for $v$ greater than some critical coupling $v_1,$ or the entire spectrum of $H$ is discrete for $v>0.$  The kinetic potential $\bar{f}(s)$ associated with the potential shape $f(r)$ is defined (for the ground state) by
a constrained minimization in which the mean kinetic energy $s = \langle-\Delta\rangle$ is kept constant:
\begin{equation}\label{defKP}
\bar{f}(s) = \inf_{{{\scriptstyle \psi \in {\cal D}(H)} \atop {\scriptstyle \|\psi\| = 1}} \atop {\scriptstyle (\psi, -\Delta\psi) = s}} (\psi, f\psi).
\end{equation}
The eigenvalue $F(v)$ of $H$ is then recovered from $\bar{f}(s)$ by a final minimization over $s$:
\begin{equation}\label{efroms}
F(v) = \min_{s>0}\left[s + v\bar{f}(s)\right].
\end{equation}
The spectral function $F(v)$ is concave ($F''(v) < 0$); moreover, we have shown \cite{inv1} that
\begin{equation}\label{convexities}
F''(v)\bar{f}''(s) = -\frac{1}{v^3}.
\end{equation}
Thus $F(v)$ and $\bar{f}(s)$ have opposite convexities and are related by the following Legendre
 transformations $\bar{f}\leftrightarrow F$ \cite{GF}:
\begin{eqnarray}
\bar{f}(s) = F'(v),\quad s = F(v)-vF'(v)\label{legendre1},\\
1/v = - \bar{f}'(s),\quad F(v)/v = \bar{f}(s) - s\bar{f}'(s)\label{legendre2}.
\end{eqnarray}
$F(v)$ is not necessarily monotone, but $\bar{f}(s)$ is monotone decreasing.  \eq{legendre1} enables us also to work with the coupling  as a minimization parameter. For this purpose we write the coupling as $u$ and we have from \eq{efroms}
\begin{equation}\label{efromu}
F(v) = \min_{u>0}\left[F(u)-uF'(u)  + vF'(u)\right].
\end{equation}
This is particularly useful in cases where $\bar{f}(s)$ is difficult to find explicitly.

\medskip

 By considering finite-dimensional linear spaces in ${\cal D}(H)$ we can 
extend these definitions and transformations \cite{env3,env6} to apply to the excited states. For example, in $d=3$ dimensions, if $D_n$ is an $n$-dimensional linear space of radial functions $\{\phi_i\}$ contained  in ${\cal D}(H)$ and in the angular-momentum space labelled by $Y_{\ell}^m$, then we have
\begin{equation}\label{enL}
E_{n\ell} = \inf_{D_n}\,\sup_{{{\scriptstyle \psi \in D_n} \atop {\scriptstyle \|\psi\| = 1}}} (\psi, H\psi).
\end{equation}
We now scale the linear space $D_n$ so that we can fix $\langle-\Delta\rangle = s.$  We define a scaling operator $\hat{\sigma}$ by $(\hat{\sigma}\psi)(r) = \psi(\sigma r),\,\sigma > 0$ Then, if we let
\begin{equation}\label{sigmaD,}
\hat{\sigma}D_n = {\rm span}\{\hat{\sigma}\phi_i\}_{i= 1}^n,
\end{equation}
we may then define 
\begin{equation}\label{usigmaDn}
{\mathscr D}_n = \bigcup_{\sigma > 0}\hat{\sigma}D_n.
\end{equation}
We note that this union of linear spaces is not itself a linear space. We may now define the excited-state kinetic potentials by
\begin{equation}\label{kpn}
\bar{f}_{n\ell}(s) =  \inf_{{\mathscr D}_n}\,\sup_{{{\scriptstyle \psi \in {\mathscr D}_n} \atop {{\scriptstyle \|\psi\|=1 }\atop{ (\psi,-\Delta\psi) = s}}}} (\psi, f\psi).
\end{equation}

 We shall not usually need to use this abstract definition since \eq{legendre1} enables us to generate kinetic potentials directly from known energy functions $F(v).$  For example, 
 with the s-states for the Hulth\'en potential $f(r) = -1/(e^r-1)$, we have from \eq{ehulthen} 
\begin{equation}\label{shulthen}
\bar{f}_n(s) =-\half\left[\left[\frac{4s}{n^2}+1\right]^{\half}-1\right],\quad n = 1,2,3,\dots .
\end{equation}

The point of these alternative expressions for the spectral curves will become clear when we discuss approximations in the next section. Another form of expression, useful for our present task, is obtained if we change the kinetic-energy parameter from $s$ to $r$ itself by inverting the (monotone) function $\bar{f}(s)$ to define the associated 
$K$-function by
\begin{equation}\label{Kfunction}
K^{[f]}(r) = s = \left(\bar{f}^{-1}\circ f\right)(r).
\end{equation}
Now the energy formula Eq.\,(\ref{efroms}) becomes
\begin{equation}\label{efromr}
F(v) = \min_{r>0}\left[K^{[f]}(r)  + vf(r)\right].
\end{equation}
A sleight of hand may be perceived here since $K$ depends on $f$.  However,  we do now have a relation that has $F$ on one side and $f$ on the other: our goal is to invert this expression, to effect $F\rightarrow f.$ We shall do this in section~(4) by using a sequence of approximate $K$-functions which do not depend  on $f$.
\medskip

Another class of soluble problems is the set of pure-power potentials the form of  whose energy functions are  determined by scaling arguments. We have \cite{env3}
\begin{equation}\label{powerpots}
f(r) = \sgn(q)\,r^q\rightarrow F^{(q)}_{n\ell}(v) = E_{n\ell}(q)\,v^{\frac{2}{2+q}},\quad q > -2, \quad q\ne 0,
\end{equation}
where the unit-coupling eigenvalues $E_{n\ell}(q)$ are known in special cases, such as 
the Coulomb problem $E_{n\ell}(-1) = -1/(4(n+\ell)^2) $ and the harmonic oscillator 
$E_{n\ell}(2) = 4n + 2\ell -1,$ where $n = 1,2,3,\dots$ and $\ell = 0,1,2,\dots$    The corresponding kinetic potentials for this family become from Eqs.\,(\ref{legendre1}) and (\ref{powerpots})
\begin{equation}\label{powerkp}
\bar{f}(s) = \frac{2}{q}\left|\frac{qE_{n\ell}(q)}{2+q}\right|^{\frac{q+2}{2}}\,s^{-\frac{q}{2}}.
\end{equation}
By contrast, the $K$-functions are much simpler and are given by
\begin{equation}\label{powerk}
K^{[q]}(r) = \frac{P_{n\ell}^2(q)}{r^2},
\end{equation} 
where
\begin{equation}\label{powerp}
P_{n\ell}(q) = \left|E_{n\ell}(q)\right|^{\frac{2+q}{2q}}\left[\frac{2}{2+q}\right]^{\frac{1}{q}}\left|\frac{q}{2+q}\right|^{\half},\quad q\ne 0.
\end{equation}
For example
\begin{equation}\label{Pm12}
P_{n\ell}(-1) = n + \ell\quad {\rm and}\quad P_{n\ell}(2) = 2n + \ell -\half.
\end{equation}
By defining $P_{n\ell}(0) = \lim_{q\rightarrow0}\,P_{n\ell}(q),$ we not only make $P_{n\ell}(q)$ continuous at $q = 0$ but we also exactly accommodate the $\ln(r)$ potential \cite{env3}. It has been proved that the $P(q)$ functions are monotone increasing \cite{env4}, and they appear to be concave. By contrast, the functions $E(q)$ are more complicated and have infinite slopes at $q = 0.$ If we denote the eigenvalues of $-\Delta + \ln(r)$ by $E^L_{n\ell},$ we have by scaling
\begin{equation}\label{logpot}
f(r) = \ln(r) \rightarrow F_{n\ell}(v) = -\half v\ln(v/v_{n\ell}),\quad v_{n\ell} = \exp(2E^L_{n\ell}).
\end{equation}
Meanwhile, the corresponding  kinetic potential and $K$-function are given by
\begin{equation}\label{logk}
\bar{f}(s) = E_{n\ell}^L -\half \ln(2es)\quad{\rm and}\quad K^{[\ln]}(r) = \frac{P_{n\ell}^2(0)}{r^2}.
\end{equation}
Numerical values for $P_{n\ell}(q)$ may be found for $d = 3$ in Ref.\,\cite{hallp} and for more general $d >1$ in Ref.\,\cite{hallq}.
\medskip

The remaining task for this section is to note the scaling and invariance properties of $\bar{f}(s)$ and $K^{[f]}(r).$
We recall that, for a given eigenvalue, we recover the energy by the expressions
\[
F(v) = \min_{s>0}\left[s + v\bar{f}(s)\right] = \min_{r >0}\left[K^{[f]}(r) + vf(r)\right].
\]
It follows that if we scale and shift $f(r)$ to $A\,f(r/b) + B,$ $a,\,b >0,$ then the new $\bar{f}$ and $K$-functions are given by
\begin{equation}\label{scalek}
\bar{f}(s) \rightarrow A\,\bar{f}(b^2s) + B \quad {\rm and}\quad K^{[f]}(r)\rightarrow \frac{1}{b^2}K^{[f]}(r/b).
\end{equation}
The $K$-function has strong invariance since it must be employed with a changed $f(r)$ and the impacts of the parameters $A$ and $B$ are not overlooked. For pure powers and the log potential, the $K$-function is completely invariant with respect to this family of potential-shape transformations, including scale.  We shall show in  the next section that if $f(r) = g(h(r)),$ and $g$ is convex, then the approximation $K^{[f]}\approx K^{[h]}$ leads to energy lower bounds; it also removes $f$ from $K$ and suggests that inversion $F\rightarrow f$ might be possible via \eq{efromr}. When $g$ is concave, we obtain upper bounds.

\section{Smooth transformations and envelope approximations}
In this section  we consider potential shapes $f(r)$ that may be written as smooth transformations $f(r) = g(h(r))$ of a `basis potential' $h(r)$.  The idea is that we know the spectrum of $-\Delta +vh(r)$ and we try to use this to study the spectrum of $-\Delta + vf(r).$ When the transformation function $g$ has definite convexity ($g''$ does not change sign), the kinetic-potential formalism immediately allows us to determine energy bounds. This is a consequence of Jensen's inequality \cite{Jensen}, which may be expressed in our context by the following:
\begin{eqnarray}\label{jineq}
\nonumber{\rm g~is~convex}~ &(g''\ge 0) \Rightarrow (\psi, g(h)\psi) \ge g((\psi,h\psi)), \\
{\rm g~is~concave}~ &(g''\le 0) \Rightarrow (\psi, g(h)\psi) \le g((\psi,h\psi)).
\end{eqnarray}
More specifically, we have for the kinetic potentials
\begin{equation}\label{kpineq}
g''\ge 0 \Rightarrow \bar{f}(s) \ge g(\bar{h}(s));\quad
g''\le 0 \Rightarrow \bar{f}(s) \le g(\bar{h}(s)).
\end{equation}
We can summarize these results by writing $\bar{f}(s)\approx g(\bar{h}(s))$ and remembering that $\approx$ is an inequality whenever $g$ has definite convexity. The expression of these results in terms of $K$-functions is even simpler, for we have
\begin{equation}\label{kg}
K^{[f]} = \bar{f}^{-1}\circ f \approx (g\circ\bar{h})^{-1}\,\circ(g\circ h)= 
 \bar{h}^{-1}\circ h = K^{[h]}.
\end{equation}
Thus $K^{[f]}\approx K^{[h]}$ is the approximation we sought, that no longer depends on $f.$ The corresponding energy bounds are provided by
\begin{equation}\label{eapprox}
E = F(v) \approx \min_{s > 0}\left\{s + vg\left(\bar{h}(s)\right)\right\} =
 \min_{r>0}\left\{K^{(h)}(r) + vf(r)\right\}.
\end{equation}
Many examples and results derived from \eq{eapprox} have been discussed in earlier papers \cite{env1,env2,env3,env4,env5,env6}. It remains here for us to describe another equivalent but geometrical approach which originally led, without Jensen's inequality, to the term `envelope method'. This alternative is particularly useful when, as for the Dirac equation \cite{halld1,halld2}, the kinetic-potential apparatus turns out to be complicated.
\medskip

We study potentials of the form $f(r) = g(h(r))$ and we suppose that an eigenvalue of the operator $-\Delta +vh(r)$ is known and is given by $H(v).$  We now consider the `tangential potentials' given by
\begin{equation}\label{tpots}
f^{(t)}(r) = a(t)h(r) + b(t), 
\end{equation}
where the coefficients are
\begin{equation}\label{ab}
a(t) = g'(h(t)), ~~{\rm and}~~ b(t) = g(h(t))-h(t)g'(h(t)).
\end{equation}
Thus the original potential $f(r)$ is the envelope of the family of tangential potentials $\{f^{(t)}(r)\}$.
Let us suppose, for definiteness, that $g(h)$ is convex.  Then each tangential potential is of the form $ah(r)+b$ lying below $f(r)$ and whose eigenvalues are known in terms of $H(v)$. Thus, the eigenvalue $F(v)$ of $-\Delta + vf(r)$ is bounded below by that of the `best' lower tangential potential, the envelope of the family of  lower energy curves. Explicitly, we have
\begin{equation}\label{elower2}
F(v) \ge \max_{t > 0}\left[H(a(t)v) + b(t)v\right].
\end{equation}
If we use the expressions for $a(t)$ and $b(t)$ given by \eq{ab}, then we find that the critical point satisfies $H'(u) = h,$ where $u = vg'(h)$. Meanwhile, the critical value is given by $H(u)-uH'(u) + vg(H'(u)).$  But this is exactly what we obtain from the expressions $s = H(u)-uH'(u),$ $\bar{h}(s) = H'(u),$ and $\bar{f}(s)\approx g(\bar{h}(s))$, that is to say
\begin{equation}\label{efromgu}
F(v) \approx \min_{u>0}\left[H(u)-uH'(u)  + vg\left(H'(u)\right)\right].
\end{equation}
This corresponds to \eq{efromu}, with the envelope approximation for the kinetic potential.

\section{The envelope inversion sequence}
We suppose that an eigenvalue $E$ of $H = -\Delta +vf(r)$ is known as function $E = F(v)$ of the coupling parameter $v > 0.$ In some cases, such as the square well, the discrete eigenvalue may exist only for sufficiently large coupling,  $v > v_1.$ The kinetic potential $\bar{f}(s)$ may be obtained by inverting the Legendre transformation \eq{legendre1}. Thus
\begin{equation}\label{fbarfromF}
F(v) = \min_{s>0}\left[s+v\bar{f}(s)\right]\rightarrow \bar{f}(s) = \max_{v>v_1}\left[\frac{F(v)}{v}-\frac{s}{v}\right].
\end{equation}
We shall also need to invert the relation \eq{efromr} between $F^{[n]}$ and $K^{[n]}$ by means of
\begin{equation}\label{KfromF}
K(r) = \max_{v>v_1}\left[F(v) - vf(r)\right].
\end{equation}
We begin with a seed potential $f^{[0]}(r)$ from which we  generate a sequence $\{f^{[n]}(r)\}_{n=0}^{\infty}$ of improving potential approximations.  The idea behind this sequence is that we search for a transformation $g$ so that $g(f^{[n]}(r))$ is close to $f(r)$ in the sense that the eigenvalue generated  is  close to $F(v).$ The envelope approximation is used at each stage. The best transformation $g^{[n]}$ at stage $n$ is given by using the current potential approximation $f^{[n]}(r)$ as an envelope basis.  We have:
\[
\bar{f} = g^{[n]}\circ\bar{f}^{[n]}\Rightarrow g^{[n]}=\bar{f}\circ\bar{f}^{[n]\,-1}.
\]
Thus
\[
f^{[n+1]} = g^{[n]}\circ f^{[n]} =\bar{f}\circ K^{[n]}.
\]
The resulting inversion algorithm may be summarized by the following:
\medskip

\noindent{\bf inversion algorithm}
\begin{eqnarray}
f^{[n]}(r)\rightarrow F^{[n]}(v) \rightarrow &~~K^{[n]}(r)&= \max_{u > v_1}\left[F^{[n]}(u) - u f^{[n]}(r)\right]\label{algor1},\\
&f^{[n+1]}(r)&= \max_{v > v_1}\left[\frac{F(v)}{v}-\frac{K^{[n]}(r)}{v}\right].\label{algor2}
\end{eqnarray}
The step $f^{[n]}(r)\rightarrow F^{[n]}(v)$ is effected by solving $(-\Delta + vf^{[n]})\psi = E\psi$ numerically for $E = F^{[n]}(v).$

\section{The inversion of spectral data from pure powers and the log potential}
Suppose that the given spectral function $F(v)$ derives from a pure-power potential shape $f(r) = \sgn(q)r^q.$ Then according to \eq{powerkp} we  know implicitly that the kinetic potential has the form $\bar{f}(s) = A(q)\,\sgn(q)\,s^{-q/2}.$  If we effect the inversion by starting, for example, with a Coulomb seed $f^{[0]}(r) = -1/r,$ we have $K^{[0]}(r) = P(-1)^2/r^2.$ It follows from the  inversion algorithm that
\[
f^{[1]}(r) = \bar{f}(K^{[0]}(r)) = \frac{A(q)}{(P(-1))^q}\,\sgn(q)\,r^q.
\]
Hence, by scaling \eq{scalek}, we find $K^{[1]}(r) = P(q)^2/r^2,$ and, in one more step, we have
\[
f^{[2]}(r) = \bar{f}(K^{[1]}(r)) = \sgn(q)\,r^q = f(r).
\]
\medskip

Similarly, if $F(v)$ is generated by the log potential $f(r) = \ln(r),$ we have $F(v) = vF(1) -\half v\,\ln(v).$ In this case, $\bar{f}(s) = F(1) - \half\ln(2es).$  We again choose a Coulomb seed $f^{[0]}(r) = -1/r,$ for which $K^{[0]}(r) = P(-1)^2/r^2.$ It follows from the  inversion algorithm that
\begin{eqnarray}
f^{[1]}(r) = \bar{f}(K^{[0]}(r)) &=& F(1) - \half\ln\left(2eP(-1)^2/r^2\right)\nonumber \\
 &=& F(1) -\half\ln\left(2eP(-1)^2\right) + \ln(r).\nonumber
\end{eqnarray}
By scaling \eq{scalek} we see that $K^{[1]}(r) = K^{[\ln]}(r) = P(0)^2/r^2,$ and  it follows that
\[
f^{[2]}(r) = \bar{f}(K^{[1]}(r)) = \ln(r).
\] 
Any pure-power (or the log) seed would yield the  same exact inversion in two steps for these cases.
\section{Inversion of Hulth\'en spectral functions}
The Hulth\'en potential has shape $f(r) = -1/(e^r-1)$ and $s$-state eigenvalues $F_n(v)$ given in \eq{ehulthen}. This  potential has a Coulomb-like singularity at the origin as evidenced by the Coulomb-like behaviour  
$F_n(v) \sim -c\,v^2$ of the spectral curves for large $v.$ We therefore use a Coulomb seed $f^{[0]}(r) = -1/r$ to begin the  inversion of $F(v) = F_1(v).$  As before, we have from this seed that $K^{[0]}(r) = P(-1)^2/r^2.$  The first inversion step can be made analytically.  By using $\bar{f}(s)$ from \eq{shulthen} with $n=1$ we obtain
\begin{equation}\label{f1hulthen}
f^{[1]}(r) = \bar{f}(K^{[0]}(r)) = -\half\left[\left[\frac{4}{r^2}+ 1\right]^{\half} -1\right].
\end{equation}
This step is shown in Fig.\,(1) as  the curve clearly between the seed and the `goal'.  The iteration is repeated twice more numerically and the graph shows $f^{[3]}(r)$, which almost coincides with  the  goal $f(r).$  We have found similar convergence when inverting excited-state spectral curves $F_n(v)$ for $n>1$  by the same algorithm. The convergence is better in these cases if the seed $K$-function $K^{[0]}(r) = (P_{n\ell}(-1)/r)^2$ corresponds to the  same radial excitation $P_{n\ell}(-1) = n+\ell = n$ as that of the energy curve that  is being inverted; with this choice, the first iteration $f^{[1]}(r)$ is {\it exactly} given by \eq{f1hulthen} for all $n\ge 1.$
\begin{figure}[!h]
\centering
\includegraphics[height=12cm,width=15cm]{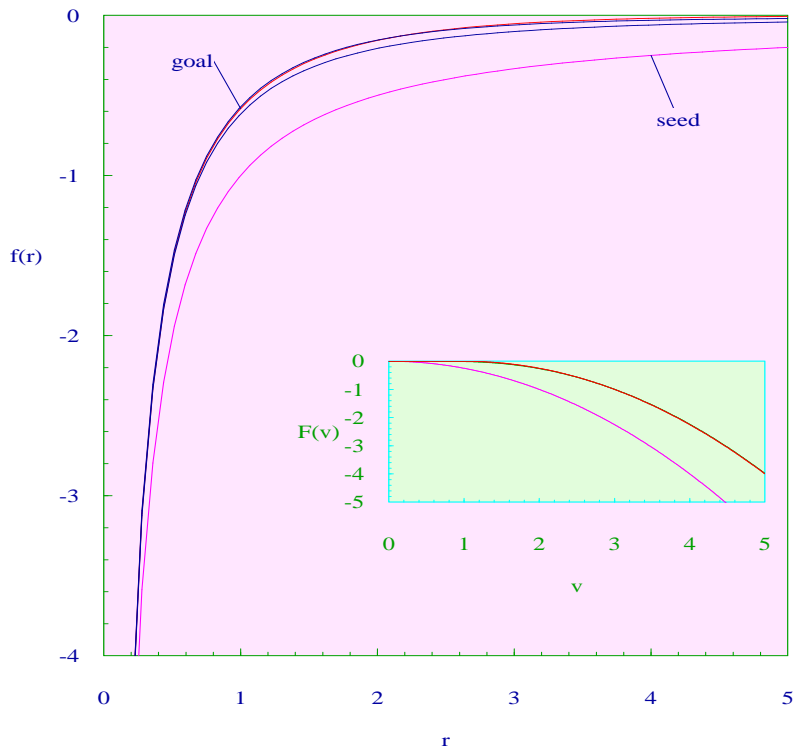}
\caption{The inversion of Hulth\'en spectral data $F(v)= -(v-1)^2/4, v \ge 1,$ shown as the upper inset curve; the lower inset curve is the Coulomb spectral function $-v^2/4.$  In the main graphs the Coulomb seed $f^{[0]}(r) = -1/r$ is used to start the inversion  iteration sequence. The diagram also shows the first iteration $f^{[1]}(r)$ and the third iteration $f^{[3]}(r)$, which almost coincides with the Hulth\'en goal $f(r) = -1/(e^r-1).$}\label{fig1}
\end{figure}

\section{Inversion of spectral data from potentials  of the form $f(r) = -a/r + b\,w(r),$ where $w(r)=\sgn(q)r^q$ or $\ln(r).$}
\begin{figure}
\centering
\includegraphics[height=12cm,width=15cm]{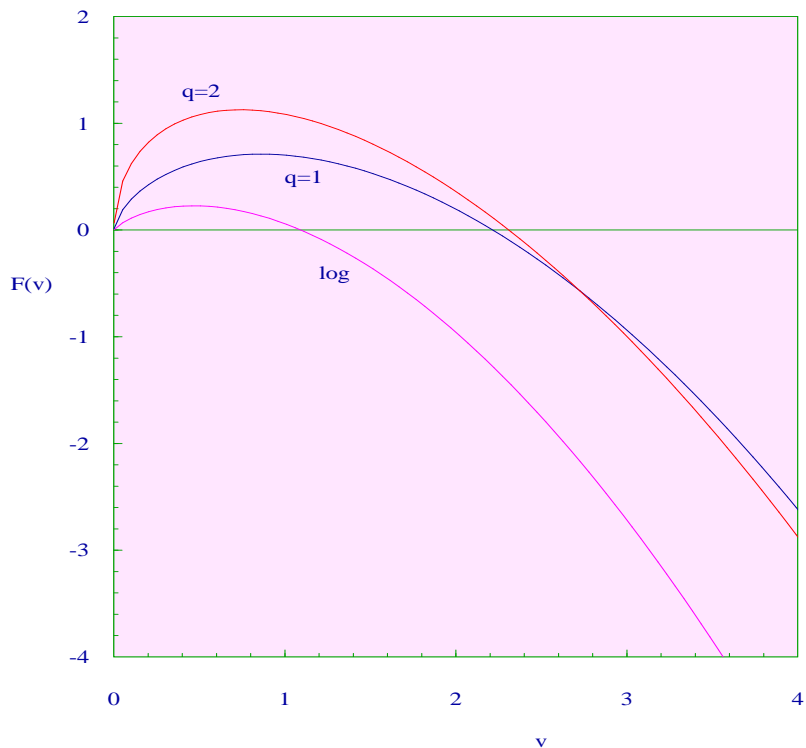}
\caption{The ground-state eigenvalue curves $F(v)$ for the operator $H = -\Delta + v(-1/r + \half w(r)),$ where $w(r)$ is respectively $r$, $r^2,$ and $\ln(r).$ The inversion of these spectral curves are shown respectively in Figs.~(3 - 5).}\label{fig2}
\end{figure}

\begin{figure}
\centering
\includegraphics[height=12cm,width=15cm]{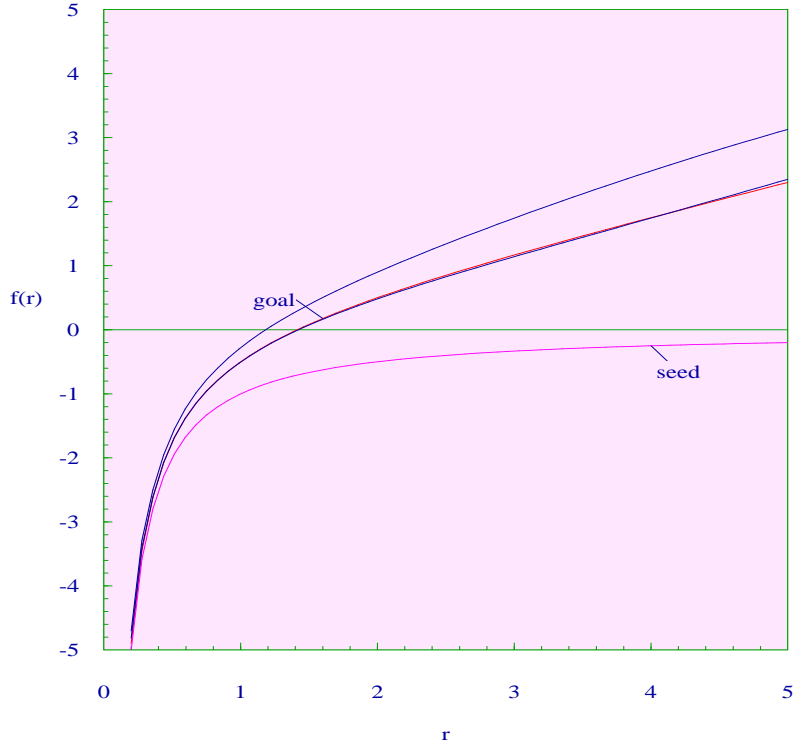}
\caption{The inversion of the spectral data shown in Fig.~(2) for the Coulomb-plus-linear potential $f(r) = -1/r + \half r.$ The diagram shows the seed $f^{[0]}(r) = -1/r,$ the first iteration $f^{[1]}(r),$ and the third iteration $f^{[3]}(r),$ which almost coincides with the goal, the exact potential $f(r)$.}\label{fig3}
\end{figure}

\begin{figure}
\centering
\includegraphics[height=12cm,width=15cm]{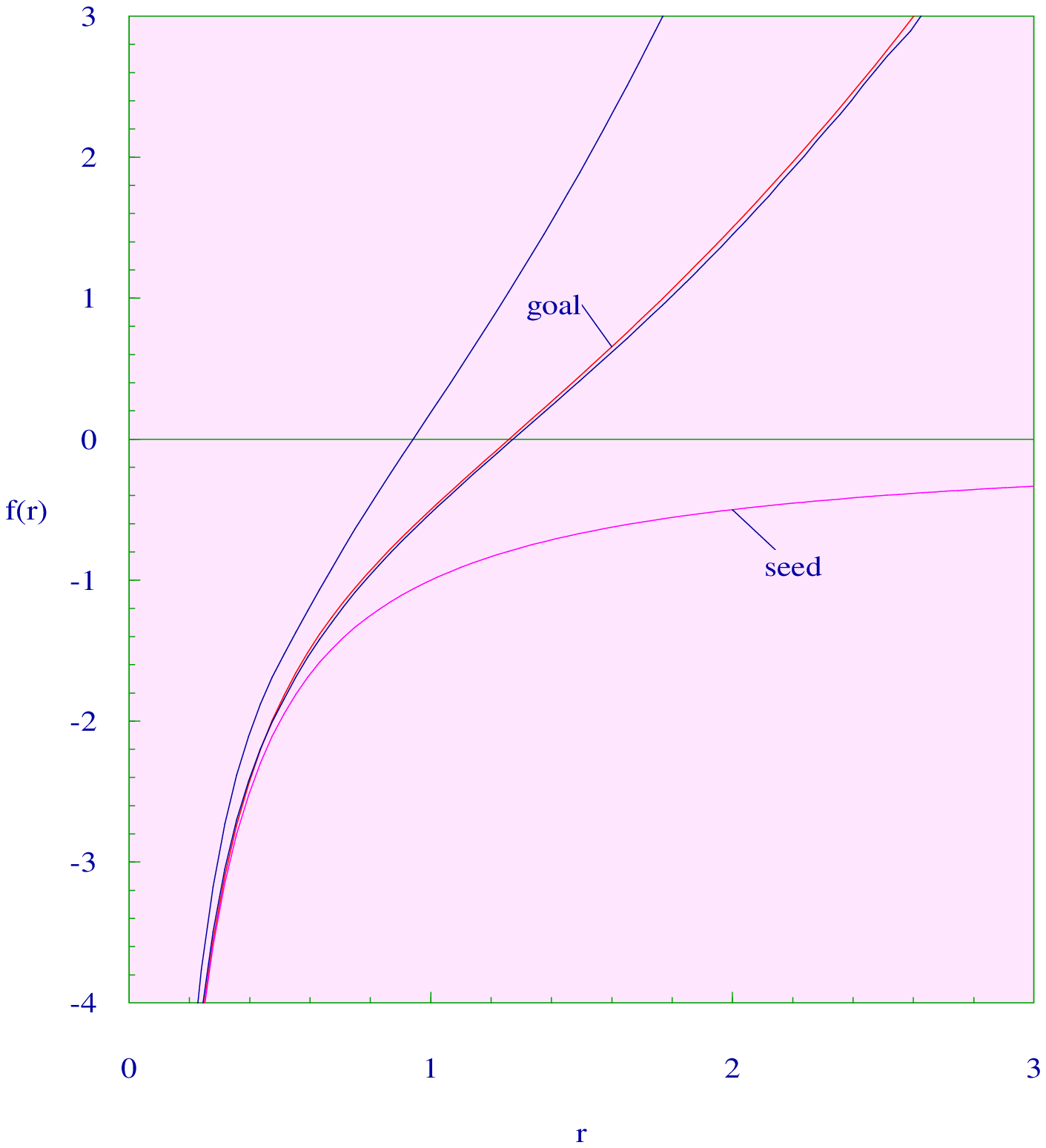}
\caption{The inversion of the spectral data shown in Fig.~(2) for the Coulomb-plus-oscillator potential $f(r) = -1/r + \half r^2.$ The diagram shows the seed $f^{[0]}(r) = -1/r,$ the first iteration $f^{[1]}(r),$ and the third iteration $f^{[3]}(r),$ which almost coincides with the goal, the exact potential $f(r)$.}\label{fig4}
\end{figure}

\begin{figure}
\centering
\includegraphics[height=12cm,width=15cm]{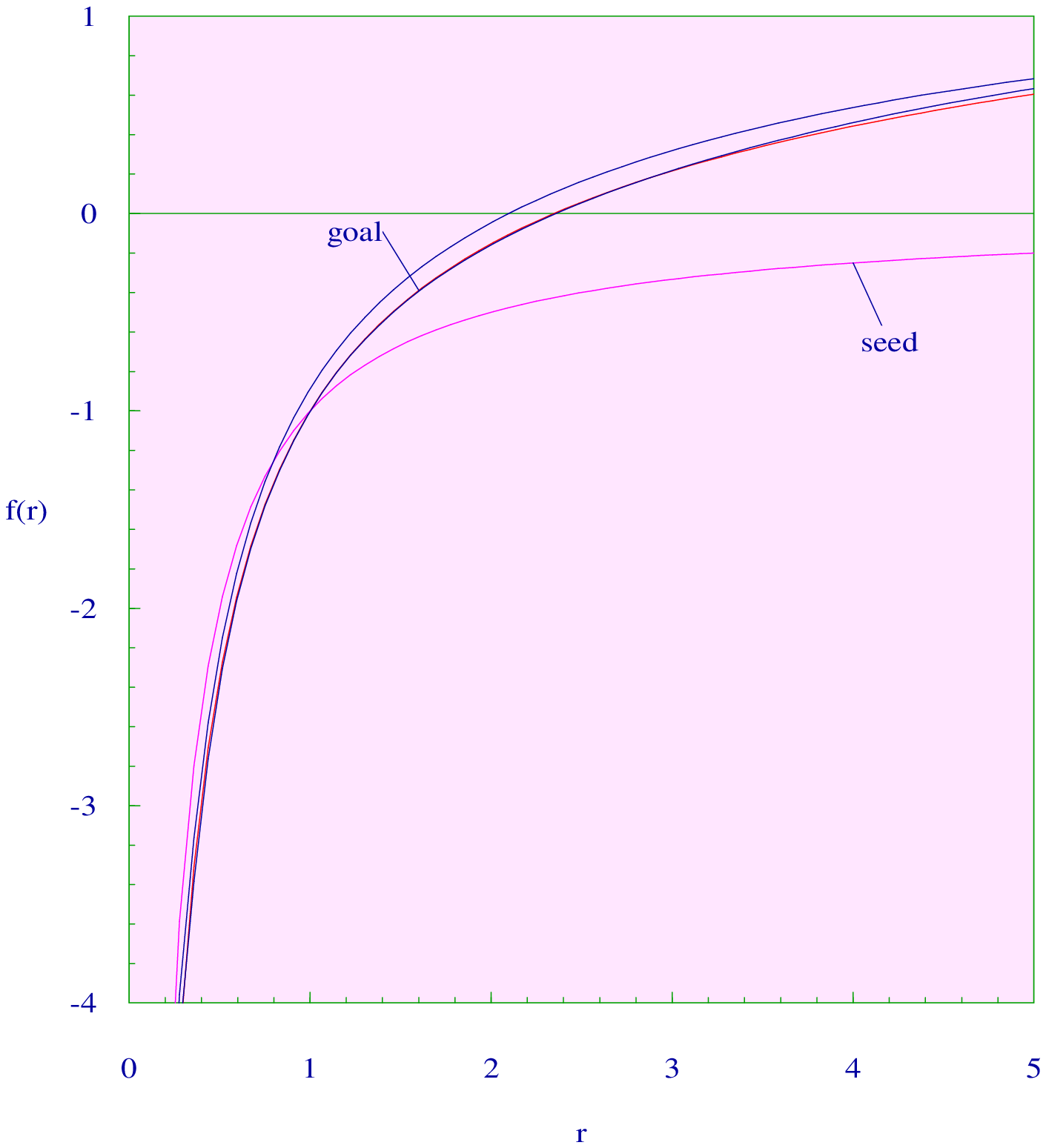}
\caption{The inversion of the spectral data shown in Fig.~(2) for the Coulomb-plus-log potential $f(r) = -1/r + \half \ln(r).$ The diagram shows the seed $f^{[0]}(r) = -1/r,$ the first iteration $f^{[1]}(r),$ and the third iteration $f^{[3]}(r),$ which almost coincides with the goal, the exact potential $f(r)$.}\label{fig5}
\end{figure}

We now consider three examples of spectral data arising from singular potentials of the form
\begin{equation}\label{couw}
f(r) = -\frac{a}{r} + b\,w(r),\,{\rm where}\, w(r) = r,\,r^2,\,\ln(r).
\end{equation}
Graphs of the corresponding ground-state energy functions $F(v)$ for $a =1$ and $b = \half$ are shown in Fig.\,(\ref{fig2}).
The inversion algorithm of section\,(4) has been applied to each of these spectral curves and the results are shown in Figs.\,(\ref{fig3}-\ref{fig5}).  In each case the seed potential was the Coulomb potentinal $f^{[0]}(r) = -1/r.$ The curve lying above the goal is the first iteration $f^{[1]}(r);$ the third iteration is also shown and is in each case  very close to the goal of the example, namely $f(r)$ itself.
\section{Uniqueness}
An important issue concerning the inversion $F\rightarrow f$ is whether there is more than one inverse $f$ for a given $F.$   We tackled this question earlier, in Ref. \cite{inv2} $\S 6,$  for bounded symmetric potentials in $d=1$ dimension that were monotone non-decreasing
on the half-line $x >0$: under these conditions, the inverse was shown to be unique. It is straightforward to prove a similar theorem for bounded potentials in three dimensions. However, a new theorem is needed for singular potentials.  We suppose in $d=3$ dimensions that the potential shape $f(r)$ is attractive, spherically symmetric, and Coulombic for small $r$; more specifically, weshall assume that $f(r)$ can be written in the form
\begin{equation}\label{formf}
f(r) = \frac{g(r)}{r},\quad{\rm where}\quad g(0) < 0,\, g'(r) \ge 0,
\end{equation}
and $g(r)$ is  not constant. Examples of this class of singular potential shapes are Yukawa $g(r) = -e^{-ar},$ Hulth\'en $g(r) = -r/(e^{ar} -1),$ and linear-plus-Coulomb $g(r) = -a + br^2,$ with $a,b > 0.$ We consider the ground state $\psi(r,v)$ which we suppose to be a normalized radial function so that 
\begin{equation}\label{norm}
\int_0^{\infty}\psi^{2}(r,v)r^2 dr = 1.
\end{equation}
With these assumptions, we shall prove the following

\nl
\noindent {\bf Theorem~1} ~~~{\it The potential shape $f(r)$ in $H = -\Delta + vf(r)$ is uniquely determined by the ground-state energy function $E = F(v).$}

\nl
{\bf Proof}
\nl
For this singular class of potentials it is helpful to consider an additional probability density on $[0,\infty)$
 of the form $c\psi^2(r,v)r$ and to define the integrals
\begin{equation}\label{qint}
q(a,v) = \int_0^a\psi^2(r,v)rdr\quad{\rm and}\quad I(v) = \int_0^{\infty}\psi^2(r,v)rdr.
\end{equation}
In this sense, $Q(a,v) = q(a,v)/I(v)$ is the probability mass on $[0,a].$ Our first task is to establish a concentration lemma to the effect that, for each fixed $a$, this probability $Q(a,v) \rightarrow 1$ as $v$ increases without limit. From \eq{legendre1} we have
\begin{equation}\label{Fprime1}
F'(v) = \int_0^{\infty}\psi^2(r,v)r^2f(r)dr = \int_0^{\infty}\psi^2(r,v)rg(r)dr \ge I(v)g(0)
\end{equation}
and
\begin{equation}\label{Fprime2}
F'(v) =  \int_0^{a}\psi^2(r,v)rg(r)dr + \int_a^{\infty}\psi^2(r,v)rg(r)dr \ge q(a,v)g(0) + (I(v)-q(a,v))g(a).
\end{equation}Thus we find for each $a$ such that $g(a) > g(0),$
\begin{equation}\label{qoverI}
1 \ge Q(a,v) = \frac{q(a,v)}{I(v)} \ge \frac{g(a)-F'(v)/I(v)}{g(a) - g(0)}.
\end{equation}
For large $v$ the potential shape is spectrally dominated by the Coulomb term $vf(r) \sim  vg(0)/r$. The corresponding wave function, eigenvalue, and $I(v)$  are given asymptotically by
\begin{equation}\label{asymptotic}
\psi(r,v) \sim c e^{vg(0)r/2}, \quad  F(v) \sim -\frac{(vg(0))^2}{4},\quad {\rm and}\quad I(v) \sim \frac{|vg(0)|}{2}.
\end{equation}
It follows that $F'(v)/I(v) \sim g(0).$ By using these asymptotic expressions in \eq{qint} and \eq{Fprime1} we arrive at the limit
\begin{equation}\label{Fplim}
\lim_{v \rightarrow \infty}\left(\frac{F'(v)}{I(v)}\right) = g(0).
\end{equation}
Thus, for each fixed $a$ such that $g(a) > g(0)$, we have the {\bf radial concentration lemma} 
\begin{equation}\label{clemma}
1 \ge Q(a,v) \ge \frac{g(a)-F'(v)/I(v)}{g(a) - g(0)}\rightarrow 1.
\end{equation}
\medskip
Now let us suppose that the two potential shapes $g(r)/r$ and $(g(r)+\gamma(r))/r$ both have the same ground-state energy function $F(v).$  We shall prove that $\gamma = 0.$ 
By considering large $v$ we see from \eq{asymptotic} that $\gamma(0) = 0.$ We consider the  two Hamiltonians $H = -\Delta +vg(r)/r$ and $H_1 = -\Delta + v(g(r)+\gamma(r))/r$ with respective lowest normalized radial eigenstates $\psi(r,v)$ and $\phi(r,v)$, $J(v) = \int_0^{\infty}\phi^2(r,v)rdr, $ and the common eigenvalue curve is $F(v).$  From the equalities
\begin{equation}\label{Fequal}
(\psi,H\psi) = F(v) = (\phi, H_1\phi)
\end{equation}
and the variational inequalities
\begin{equation}\label{Finequal}
(\psi,H_1\psi) \ge F(v) \le (\phi, H\phi)
\end{equation}
we deduce the  complimentary inequalities
\begin{equation}\label{compineq}
(\phi,\gamma(r)/r\,\phi) = \int_0^{\infty}\phi^2(r,v)\gamma(r)rdr \le 0 \le \int_0^{\infty}\psi^2(r,v)\gamma(r)rdr = (\psi,\gamma(r)/r\,\psi). 
\end{equation}
Thus
\begin{equation}\label{compineq2}
\frac{1}{J(v)}\int_0^{\infty}\phi^2(r,v)\gamma(r)rdr \le 0 \le \frac{1}{I(v)}\int_0^{\infty}\psi^2(r,v)\gamma(r)rdr. 
\end{equation}
By a sweeping process starting at $r=0$ we now show that \eq{compineq2} implies that $\gamma \equiv 0.$  We know that $\gamma(0)=0.$ Clearly, if $\gamma(r)$ does not change sign, then it must vanish identically.  Let us suppose that $\gamma(r)$ first changes sign at $r = a$; more specifically (and without loss of generality), we suppose $\gamma(r) \ge 0$ for $r\in(0,a)$, and that $\gamma(a) = 0.$  We first note that if $g(a) = g(0),$ $g(r)$ would be constant on $[0,a]$ and $\gamma(r)$ would have to vanish on $[0,a]$ in order for $(g(r)+\gamma(r))/r$ to remain  in the same potential class. Thus we understand $r = a$ to be the first point where $\gamma(r)$ is zero, after it has been positive. We now increase $v$ so that both of the following inequalities are satisfied
\begin{equation}\label{bineq}
\frac{F'(v)}{I(v)} < g(a),\quad \frac{F'(v)}{J(v)} < g(a) + \gamma(a)= g(a).
\end{equation}
In view of the radial concentration lemma, further increase in $v$ will force both integrals in \eq{compineq2} to be positive.  This cannot be unless $\gamma \equiv 0.$ This establishes Theorem~1.\qed

\section{Conclusion}
We are presented with a spectral curve, $F(v)$, which shows how a discrete eigenvalue of the Hamiltonian $H = -\Delta + vf(r)$ depends on the coupling parameter $v > 0.$ From this data, we reconstruct the underlying potential shape $f(r)$.  This `geometric spectral inversion' is effected by a functional sequence in which the unknown potential shape $f(r)$ is first regarded as  a smooth transformation $g\left(f^{[0]}(r)\right)$ of a seed potential $f^{[0]}(r)$. The paper focuses on problems that involve singular potentials for which the Coulomb seed $f^{[0]}(r) = -1/r$ is very effective. The method of potential envelopes is used to generate a sequence of approximations $\{f^{[n]}(r)\}_{n=0}^{\infty}$ for which the corresponding transformations gradually approach the  identity. At this stage of the investigation we offer an explicit heuristic inversion algorithm rather than an abstract convergence theorem. However, for the class of attractive central potentials with shapes $f(r) = g(r)/r,$ with $g(0)< 0$ and $g'(r)\ge 0,$ we prove that the ground-state energy curve $F(v)$ determines $f(r)$ uniquely. 
\section*{Acknowledgments}

One of us (RLH) gratefully acknowledges both partial financial support
of this research under Grant No.\,GP3438 from the Natural Sciences
and Engineering Research Council of Canada and the hospitality of
the Institute for High Energy Physics of the Austrian Academy of
Sciences, Vienna, where part of the work was done.

\section*{}

\end{document}